# Effect of varying preheating temperatures in electron beam powder bed fusion: Part I Assessment of the effective powder cake thermal conductivity

Gitanjali Shanbhag[1], Mihaela Vlasea[1]

[1]University of Waterloo, Waterloo, ON N2L 3G1, CANADA

**Abstract:**

One of the major barriers in adapting the existing EB-PBF process parameters to a new powder material system is controlling the preheating conditions such that every layer of powder results in enough partial sintering to create a coherent powder cake. To be able to understand the powder sintering process and adapt it to other materials, we must look at the degree of sintering and the effective thermal conductivity of the powder bed. An in-depth understanding of these characteristics will help tailor the preheating conditions and furthermore, make it easier to remove/de-powder intricate parts after build completion without compromising the advantages of the preheating phenomenon. This study evaluates the impact of preheating temperature on the in-situ powder cake properties. Three different preheat temperatures, 650 °C, 690 °C and 730 °C, are employed to a Ti-6Al-4V powder cake and in each standalone build, unique powder-capture artefacts are fabricated to be able to analyze the in-situ powder cake properties using X-ray computed tomography. An empirically-derived model for thermal conductivity of the powder cake as a result of changing the preheating temperatures, was obtained. The results demonstrated that the effective thermal conductivity of the powder cake at a given preheating temperature strongly depends on the packing density, contact size ratio and coordination number. An increase in preheating temperature, lead to a linear increase in the packing density (from 58.42 ± 1.10 % to 61.87 ± 0.96 %), contact size ratio (from 0.45 ± 0.004 to 0.48 ± 0.003), coordination number (from 3.36 ± 0.05 to 3.58 ± 0.05) and the effective thermal conductivity (from 1.75 ± 0.07 W/m/K to 2.11 ± 0.07 W/m/K) of the powder cake. The findings in this work will be deployed to assess and to correlate the effects of different powder conductivity and Preheat Themes on the surface topography and geometric fidelity of components with simple and complex geometries.

# 1   Introduction

The energy source in electron beam powder bed fusion (EB-PBF) processes, is used for both partial sintering (during preheating) and fusing (during melting) powder particles together [1]. The electron beam interaction with metallic powder develops a charge distribution around the build plate. If this charge exceeds a critical limit, the repulsive forces between the negatively-charged powder particles can cause particle motion and result in an avalanche effect, also known as "smoke" or "smoking effect" [2]. As a result of smoking, a powder cloud is created, which can spread detrimentally throughout the build chamber and potentially up into the electron beam gun, all within a matter of milliseconds. In order to avoid this smoking effect, the powder needs to be partially sintered [3]. This powder sintering is performed by using the Preheat Theme. There are two steps in preheating, Preheating 1 (PH1) and Preheating 2 (PH2). PH1 is required for securing the powder particles strongly such that the electron beam is able to sweep over the powder bed without the formation of any smoke. PH1 covers the entire build plate and enables the electron beam to jump between melt areas; making the electron beam jump safe. PH2 is required to facilitate the melting of powder particles in select regions corresponding to the part cross-section at that layer and to prevent swelling in parts. PH2 covers each individual melt region (as shown in Figure 1) and uses a higher energy electron beam to provide mechanical anchoring of parts and supports (if any) [4]. The PH1 and PH2 stages are followed by the melting of the powders at the location corresponding to the part slice, as controlled by the Melt Theme; the effect of the Melt Theme is out of scope for this study.

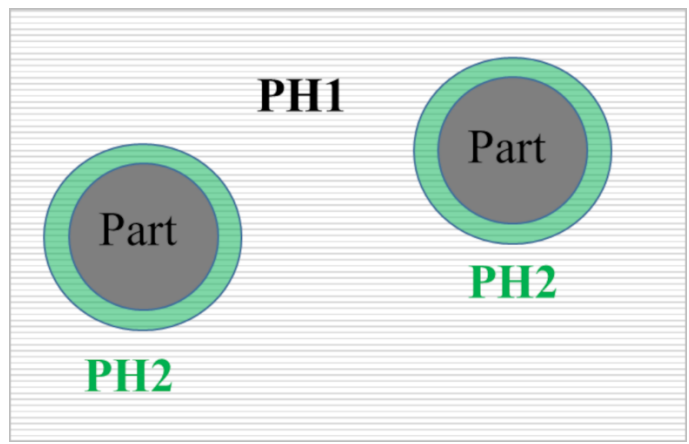

Figure 1 Illustration of where the PH1 and PH2 theme regions are in effect, with respect to a part slice within a layer. PH1 and PH2 are applied to sinter the powder cake. The Melt Theme is applied to melt the cake and create the part layer.

The reason for using a higher energy electron beam is to create a more uniform heat environment by lowering the difference in thermal conductivity between parts and powder. PH2 ensures that the electron beam is melt safe and can move on to the Melt Theme. The Preheat Theme can be controlled by varying the beam current (mA), beam speed (mm/s), number of repetitions, focus offset (mA) and scan order. The beam current controls the energy input during the heating process, the beam speed controls the exposure time of the beam, the focus offset controls the beam sharpness or beam diameter during preheating. An additional solution to the smoking issue is introducing a small partial pressure of helium into the vacuum environment. This inert gas creates $He^+$ ions above the powder bed that pick-up electrons from the powder bed and reduce the risk of smoke. However, the introduction of helium only takes place during the melting stage. In addition, He gas increases the cost of the process, thus increasing its use is generally prohibitive.

## 1.1 Preheating studies in EB-PBF of Ti-6Al-4V

Studies have shown that preheating can increase the mechanical strength, electrical and thermal conductivity of the sintered powder, as well as improve the beam-powder interaction efficiency [5]–[8]. Preheating has also been shown to reduce the formation of balling phenomena [5], [9] and lower the thermal gradient during melting, thus reducing distortion and warpage in the manufactured components [7], [10], [11]. However, Sigl et al. [12] identified a few drawbacks. They mentioned that (i) preheating increases the total build time and energy consumption, (ii) a powder recovery system (PRS) is required break up the powder cake to retrieve the final parts, and (iii) preheating may limit the small details in complex internal geometries as it is difficult to remove the partially sintered powder particles. The presence of partially sintered powder particles in intricate or complex internal geometries and features may lead to an increase surface roughness of the part as well.

Drescher et al. [3] looked into modifying PH1 to achieve a higher build efficiency. The procedure was to replace the default PH1 strategy with one that preheats only the area that is to be consecutively melted. Thus, the powder surrounding the part is not sintered as strongly and the powder cake is expected to be less cohesive. This would lead to the easy removal of partially sintered powder and a reduction of time for retrieving final parts, in turn increasing productivity. They concluded that modifying the PH1 step led to an increase in build efficiency by 23%. They reported that the mechanical properties and microstructure remained unchanged. However, the powder had an increase in oxygen content and higher brittleness. Leung et al. [13] studied the effect of applied preheat energy per unit area ($E_a$) on the microstructure, hardness and porosity of parts on an A2XX machine with a layer thickness of 70 μm for Ti-6Al-4V. The applied preheat energy ($E_a$) in the base sinter was calculated as shown in Equation 1.

$$E_a = \left(\frac{E_{line}}{L_0}\right) * N \qquad 1$$

where $E_{line}$ is the line energy expressed as (Equation 2):

$$E_{line} = \frac{accelerating\ voltage * beam\ current}{scan\ velocity} \qquad 2$$

and $L_0$ is the line offset and N is the number of scan repetitions. They observed that the thermal conductivity increases with an increase in $Ea$. Another observation was that the micro-hardness of the components reduces with increasing $Ea$ due to microstructural coarsening of Ti-6Al-4V caused by annealing. They identified that an $Ea$ value of 411 kJ m$^{-2}$ was optimum as it produced parts with a high hardness and low dimensional deviations from the computer-aided design (CAD) model.

There is a need for identifying a suitable preheating strategy in order to address the disadvantages presented by Sigl et al. [12] and influence the powder cake properties for easy powder removal after the build has been completed. In order to do this, it is important to understand the effect of preheating and sintering on the powder particles and their thermal properties. The majority of the literature in this domain focuses on finite element (FE) modeling and simulation of the preheating and melting process [1], [14]–[18] to understand the heat transfer phenomena in EB-PBF. However, Landau et al. [19] have identified some of the drawbacks of such FE models. They noted that such simulations require small time increments and when using FE models, the mesh must be sufficiently fine in order to properly simulate the melt pool shape. Therefore, modeling a complete hatching path may be too computationally expensive. Secondly, the preheating process takes place over a long time and therefore using the existing moving heat source based thermal models may not be entirely practical [19]. Hence literature on modeling usually implements preheating temperatures as an initial given condition, which is not truly representative of the thermal history of the powder cake.

One of the major barriers in adapting the existing EB-PBF process parameters to a new powder material system and particle size distribution is controlling the preheating conditions such that every layer of powder obtains enough partial sintering to create a coherent powder cake, while simultaneously ensuring that the powder cake can be reconditioned into free-flowing powder with appropriate qualities for powder reuse. To be able to understand the powder sintering process and adapt it to other materials, the degree of sintering, or more explicitly the size of the sinter necks and density of each preheated layer, must be determined to predict the thermal conductivity of the powder bed. This will help tailor the preheat theme temperature and other parameters that would make it easier to remove/de-powder parts after build completion without compromising the advantages of the preheating phenomenon.

## 1.2 Thermal conductivity studies of Ti-6Al-4V powder bed

Parts manufactured by powder-bed fusion (PBF) additive manufacturing (AM) processes, that are employed in critical sectors such as aerospace, biomedical, defense, marine and automobile require rigorous qualification steps. As such, a strong understanding of the properties of the metal powder feedstock, as well as its behaviour in the process is necessary. Specifically, researchers across the globe are modelling the heat transfer process [1], [6], [10], [14], [16], [20]–[24] in PBF to comprehend the mechanisms behind the consolidation of metal powder particles. In order to do so, it is imperative to understand the thermal properties of metal powders; information on this aspect is currently limited. Inputs for thermal properties in such models are often estimated by analytical or empirical models, or directly taken from property tables and databases.

Some studies [25]–[27] have presented analytical models for calculation of the thermal conductivity in a powder bed based on geometrical considerations of the constitutive powder material system. Gusarov et al. [25] concluded that the conductivity depends on the relative density, coordination number and the contact size of the particles. This model was created to understand the sintering and binding mechanism in the laser powder bed fusion (LPBF) process. In the work by Siu et al. [26], it was concluded that the conductivity highly depends on the contact angle between the spheres when packed in simple cubic, body centered cubic or face centered cubic configuration. Although the model is difficult to generalize for complex powder particle organization, the findings capture the level of sensitivity expected as a function of packing configuration. In addition, Slavin et al. [27] concluded that the narrow gaps between adjacent particles significantly contribute to the thermal conductivity of a particle system consisting of irregularly shaped spheroidal particles. Such modeling works are important theoretical contributions, with inherent limitations in translating findings empirically, depending on the underlying model assumptions.

A few mathematical models relating the thermal conductivity of powder beds to the porosity and the gas type around the powder particles, were proposed. As such, Sih et al. [28] derived a model for a randomly packed metallic powder bed by enhancing a few variables the Zehner-Schlünder's equation [29] to incorporate conductivity of gas, contributions from porosity, and particle contact-area ratio. Yagi et al. [30] developed a model to relate the effective conductivity of a porous material to its volume void fraction. They conclude that at low temperatures the conductivity is affected only by convection, whereas at higher temperatures radiative heat transfer also plays a major role. Thümmler et al. [31] also developed a model that emphasizes the importance of pore geometries on thermal conductivity and pointed out that it can be controlled by the amount of gas content in the pores. However, the EB-PBF process takes place under vacuum and therefore these models cannot be applied to this process.

Only a handful of studies demonstrated work on empirical measurements conducted explicitly to verify and improve thermal property information used in finite element (FE) models to simulate the heat transfer mechanism in AM processes. Rombouts et al. [32] evaluated the thermal conductivity of 316L stainless steel, iron, and copper powders of various particle shapes and size ranges using photopyroelectric measurements. The authors observed that the relative density is a critical factor for thermal conductivity of the powder bed, and it was concluded that beds of irregular particle shapes or particles with a wider size distribution were more conductive. Alkahari et al. [33] studied the effect of bulk density and particle diameter on the thermal conductivity of bimodal 316L stainless steel powder. A pulsed Yb: fiber laser was used to create a heat source on the top face of the powder bed, and measurements were taken with a thermocouple embedded below the top face. The data gathered from the thermocouple output was used to calculate the thermal conductivity. The authors observed that the thermal conductivity of metal powder increased with an increase in bulk density and particle diameter and that, not surprisingly, increased porosity led to lower thermal conductivity. Arce [4] performed a comprehensive thermophysical evaluation of Ti-6Al-4V EB-PBF parts prepared from plasma rotating electrode process (PREP) and gas atomized (GA) powders as well as the powders themselves. The powder size distribution was 45-150 µm. These experiments were conducted to provide more accurate inputs for an FE model to simulate the process. Arce used differential scanning calorimetry (DSC) to measure Cp and concluded that there was not a significant difference in measured Cp values between the EB-PBF and conventionally processed Ti-6Al-4V samples. The Cp values were observed to be between $0.5 - 0.6$ J g$^{-1}$ K$^{-1}$ measured at up to 800 °C for raw powders, as-built parts and conventional processed Ti-6Al-4V. The thermal diffusivity of powder and solid samples was measured by the laser flash method by holding the samples in sapphire containers inside the instrument. Differences in the results of the two types of powders was attributed to different particle size distributions. Thermal conductivity values were acquired and compared to cast and wrought material to show that there were no considerable differences in thermal conductivity between the EB-PBF samples and conventionally processed material (values summarized in Table 1). Cheng et al. [1], [16] calculated the thermal conductivity of both solid and powder-encapsulated Ti-6Al-4V EB-PBF parts (values given in Table 1) where the powder size distribution was 45-100 µm. The authors observed that the thermal conductivity of the powder cake is < 15% of its solid counterpart. They found powder porosity to be a critical factor in the reduction of thermal conductivity. They found a value of 2.44 W m$^{-1}$ K$^{-1}$ for a sample with 50% porosity and 10.17 W m$^{-1}$ K$^{-1}$ for solid Ti–6Al–4V at 750 ºC. They also measured (with the help of scanning electron microscopy) the diameter of the sinter neck formed after preheating; such diameter values were on the order of 1 µm to 10 µm. Smith et al. [8] studied the effect of changing process parameters such as number of beam passes and line energy on the powder density, thermal diffusivity and thermal conductivity in EB-PBF, for a similar plasma atomized (PA) powder size distribution of 45-105 µm. The authors observed that the thermal diffusivity values increased with an increase in the number of beam passes over a certain area. However, the density did not change significantly with the number of beam passes, which led to the conclusion that the thermal conductivity is solely related to a morphological change; this was confirmed by electron microscopy where an increased connectivity due to necking and partial melting between particles was observed. Increased connectivity between particles is undesirable if the powder needs to be re-sieved and reused in the process.

Table 1 Thermal conductivity values for powder (W m$^{-1}$ K$^{-1}$), as-built parts, wrought rods and for wrought Ti-6Al-4V material extracted from representative plots presented by Arce [4], Cheng et al. [1], [16] and Boivineau et al. [34]. PREP = Plasma Rotating Electrode Process and GA = Gas Atomized

| Sample type | | Temperature (in °C) | | | | | | | | |
|---|---|---|---|---|---|---|---|---|---|---|
| | | 20 | 100 | 200 | 300 | 400 | 500 | 600 | 700 | 750 |
| Powder [4] | PREP | 1 | 1 | 1 | 1 | 1 | 1 | 1.1 | 1.2 | 1.5 |
| | GA | 1 | 1 | 1 | 1 | 1 | 1 | 1.1 | 1.2 | 1.5 |
| As-built Part [4] | PREP | 7 | 8 | 8.5 | 9.5 | 10.5 | 11 | 13 | 13.9 | 14.5 |
| | GA | 7 | 8.1 | 8.6 | 10 | 11 | 12 | 13.5 | 14.3 | 15.5 |
| As-built hollow part with encapsulated powder [1], [16] | | 2.7 | - | - | - | 2.8 | - | - | - | 4 |
| As-built solid part [1], [16] | | 6.2 | - | - | - | 9 | - | - | - | 10 |
| Wrought [34] | | - | 7 | 8 | 9 | 9.5 | 10 | 12 | 13 | 13.3 |

The above-mentioned literature indicates that the thermal conductivity for a powder bed is substantially different when compared to the same material obtained in its cast or wrought form. This can be attributed to the limited contact between powder particles and is furthermore influenced by complex packing conditions. Heat transfer through conduction in a solid part is faster than in powder as the vacuum gaps between powder particles work as insulators and slow down the heat transfer [24]. The conductivities are dependent on the porosity of the powder bed and the process environment. With an increase in porosity, the powder bed emissivity increases and the effective powder bed thermal conductivity decreases. However, with the small amount of existing literature, it is challenging to predict the thermal properties of the powder in vacuum when exposed to elevated temperatures for any substantial duration [8]. Significant improvements in EB-PBF part quality outcomes and powder recovery efficiency may result from more accurate estimates of powder thermal conductivity derived from experiential measurements and evaluation of the thermal properties of the powder system. Therefore, the current work will look at constructing an empirically-derived model for thermal conductivity of the powder cake in the EB-PBF process as a result of changing the Preheat Theme.

## 1.3 Assessing the in-situ powder cake properties using powder-capture artefacts

It is important to recognize that the effective thermal conductivity of a powder bed in EB-PBF depends on the neck size between the partially sintered powder particles. The diameter of this neck controls the rate of heat transfer. Tolochko et al. [35] have observed in LPBF that a decreased neck size means low thermal conductivity and vice versa. For the current work, the goal is to examine and elucidate how the preheat temperature affects the sinter neck size, layer-wise density, and thermal conductivity. As mentioned earlier, this information will help tailor the preheat theme temperature and other parameters that would make it easier to remove/de-powder parts after build completion without compromising the advantages of the preheating phenomenon. In order to study the effect of preheating on powder sintering, one needs to capture the immobilized powder cake that is formed by the two-stage preheating.

Manufacturing a powder encapsulation container during the EB-PBF process would assist in obtaining a sample of the immobilized metal powder cake during in situ building conditions. A few studies in literature tried to capture such an immobilized powder cake. Liu et al [36] manufactured a hollow cube with an open top using LPBF. The part was manufactured such that the immobilized cake was inside the container that was being built. However, the authors did not mention the container removal, powder immobilization, nor the powder removal procedure. These procedures can have a high impact on the powder bed density measurements and therefore it is important to clearly describe them. The second disadvantage of this study is that the authors did not describe the procedure of determining the actual volume of the powder container. They computed the powder bed density value by assuming that the volume of the inner cavity of the container is equal to the nominal design volume, which has been shown to be an inaccurate assumption [37]. Jacob et al. [38] built a hollow cylinder with a closed top (which was a second component – unattached to the cylinder) using LPBF. The intent was to be able to end up with unaffected powder inside the powder encapsulation container and to infer the powder bed density via mass-volume estimates. The mass of captured powder was measured by punching a hole through the lid and removing all powder. The inner volume of specimens was evaluated by mass measurements of the empty specimen and of the specimen filled with a fluid of known density to infer the fluid-accessible volume. The drawbacks of this study were certain uncertainties such as accidental removal of powder by hole punching, remainder of powder inside while draining, surface tension effects of the measurement fluid, etc. Rogalsky et al. [37] also built a hollow cylinder but with an open top. After the build was completed, tops were closed with a plastic lid with the intent to end up with unaffected powder inside the container. This study used XCT for quantitative analysis to determine the relative powder bed density as well as an infiltration method for experimental validation. However, since this study was conducted for LPBF with highly irregular powders, powder bed denudation and surface topography introduced uncertainties to the results.

This work introduces a method to assess and investigate the behaviour of powder and powder layers during the EB-PBF process. A special powder capture artefact with an internal cavity, referred to as a capsule, was used to capture the immobilized powder cake during the EB-PBF process. This powder capture artefact design assists in protecting the powder cake in a way that is unaffected from the environment during the powder recovery stage. These artefacts are also manufactured in different locations on the build plate to measure the consistency of the powder bed condition. Furthermore, the assessment of the powder cake will assist in constructing an empirically-derived model for thermal conductivity of the powder cake in the EB-PBF process as a result of changing the Preheat Theme. The findings in this work will be deployed to assess and to correlate the effects of different powder conductivity and Preheat Themes on the surface topography and geometric fidelity of components with simple and complex geometries.

## 2 Materials and Methods

### 2.1 CAD file preparation, specimen design, and build layout

SolidWorks (Dassault Systèmes, France) was used for designing and obtaining the specimen STL files. File preparation for manufacturing was then performed using Materialise Magics (Materialise, Belgium). The software was used for rescaling, positioning on a start plate, and support structure creation, where required. Slicing the files was then executed by Build Assembler (Arcam plug-in for Materialise Magics), which converts the information into an Arcam build file (.abf) that is imported to the machine.

Figure 2(b) shows a CAD design of the powder-capture artefact. Inside the artefact, a small capsule was designed where powder was captured (Figure 2(a)). Although two artefacts were manufactured (one stacked on top of the other as shown in (Figure 2(c)), XCT was performed only for the bottom artefacts.

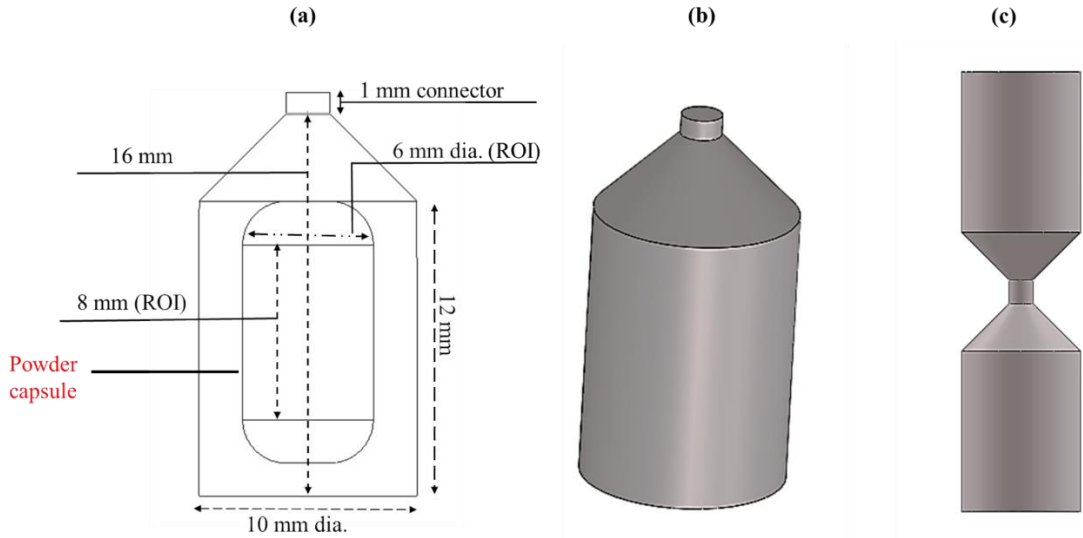

Figure 2 (a) Wireframe diagram of the powder-capture artefact (b) Solid view of the cylindrical powder-capture artefact (c) 2 artefacts stacked and connected by a connecter in the middle

The two stacked powder-capture artefacts are connected by a small 1 mm cylindrical connector. In the event that the two artefacts need to be separated, the connector can be easily cut or broken off. Note that, the artefact placed on top, appears upside down. This does not affect any analysis since the shape and orientation of the Region of Interest (ROI) remains the same with respect to the manufacturing coordinate system and the internal capsule design remains unchanged. Alongside every powder-capture artefact, there was a 15x15x30 mm part-quality artefact which will be used, in a future study, to assess the surface topography. The position of the samples on the start plate of the machine and the naming strategy are shown in Figure 3(a). All samples have their longitudinal axis perpendicular to the build platform (parallel to the build direction / Z axis). A total of 9 powder-capture artefacts and 9 part-quality artefacts were manufactured. Figure 3(b) shows a modeler view of Materialise Magics – depicting all the specimens on the build plate.

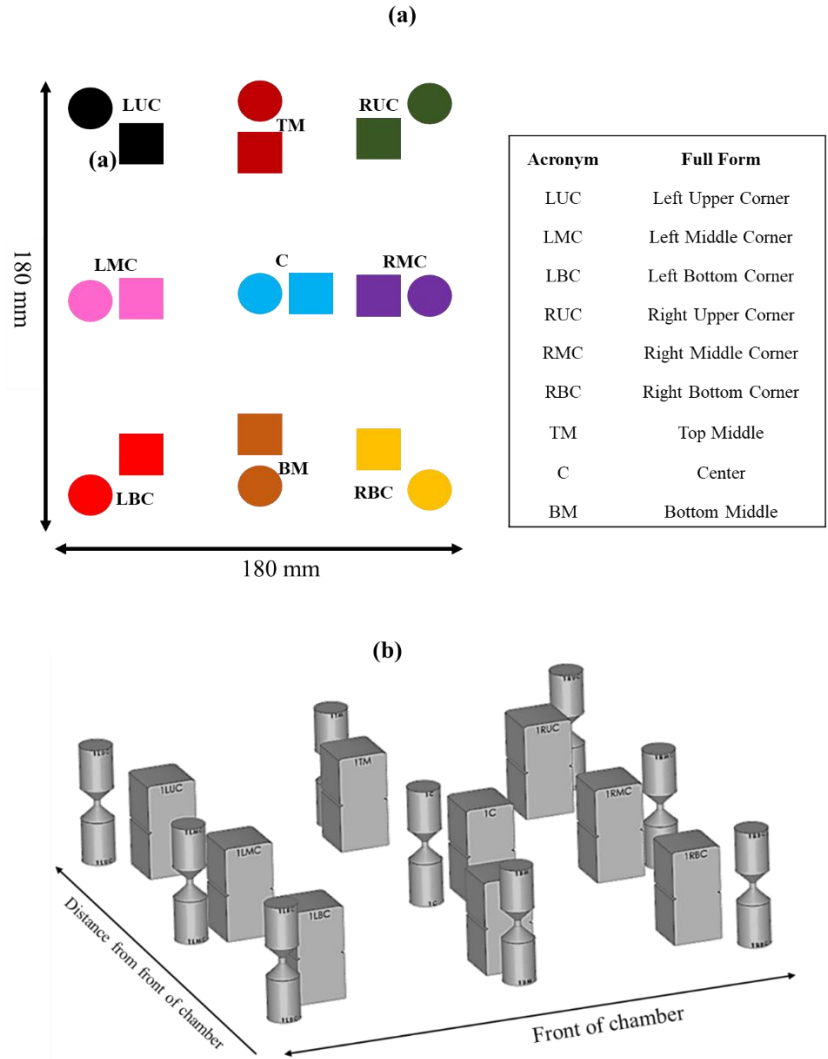

Figure 3 (a) Top view of the build plate depicting powder-capture artefacts and part-quality artefacts (left) and Labeling strategy for all specimens based on their location on the build plate (right table) (b) Modeler view of Materialise Magics showing powder-capture artefacts and part-quality artefacts on the build plate

## 2.2 Additive manufacturing of samples

All samples were produced on an Arcam A2X (Arcam, GE Additive, Sweden) EB-PBF machine. The Ti-6Al-4V powder feedstock was supplied by Advanced Coatings & Processes (AP&C, Canada) and consisted of pre-alloyed, PA powder with a size distribution of 45 μm to 105 μm. The layer thickness was set to 50 μm. The samples were built in accordance with the equipment manufacturer's default parameter settings (version 5.2.40), with only modifications being the Preheat Theme. As mentioned earlier, reheating comprises of two stages: Preheat 1 and Preheat 2. In Preheat 1, the electron beam lightly sinters the powder particles over the entire build area and in Preheat 2 the electron beam locally sinters the area where parts are supposed to be built. In the Arcam A2X model, the preheating temperature is monitored by a K-type thermocouple that is attached to the bottom of the build plate. The default preheat temperature for Ti-6Al-4V is set to 730 °C by the manufacturer. A total of three (3) separate builds were manufactured. The build height for all experiments was 37 mm. The only difference between the builds was the preheat temperature. The current study looks at three different preheating temperatures to understand the change in the in-situ properties of the powder cake. Table 2 shows the preheat parameters used for each of these experiments. The line energy, $E_{line}$ is expressed per Equation 2.

Table 2 Preheat theme process parameters

| Preheat Parameters | **Experiment 1** | | **Experiment 2** | | **Experiment 3** | |
| --- | --- | --- | --- | --- | --- | --- |
| | Preheat 1 | Preheat 2 | Preheat 1 | Preheat 2 | Preheat 1 | Preheat 2 |
| Preheat temperature (°C) | 650 | | 690 | | 730 | |
| Accelerating Voltage, $U_e$ (kV) | 60 | 60 | 60 | 60 | 60 | 60 |
| Beam current, $I_b$ (mA) | 30 | 38 | 30 | 38 | 30 | 38 |
| Scan velocity, $V_{sc}$ (ms$^{-1}$) | 10 | 13 | 10 | 13 | 10 | 13 |
| No. of scan repetitions | 2 | 3 | 2 | 3 | 2 | 3 |
| Line energy, $E_{line}$ (J.m$^{-1}$) | 180 | 175.38 | 180 | 175.38 | 180 | 175.38 |
| Line offset, $L_0$ (mm) | 1.2 | 1.2 | 1.2 | 1.2 | 1.2 | 1.2 |

Al-Bermani et al. [39] observed that a coarser microstructure (leading to a decrease in mechanical properties) can be formed if the preheating temperature is above 951 K (or 677.85 °C). Therefore, it is important to investigate if the preheating temperature can be decreased from the default preheating temperature (i.e., 730 °C).

## 2.3 X-ray computed tomography (XCT)

The XCT-scanner (Xradia 520 Versa, Zeiss, Pleasanton, CA) was set to operate at 100 kV, 90 μA, and resolution of 6 μm. The acquisition files were obtained at 1001 projections (per 360 deg of rotation). Each artefact was first scanned, capturing the powder capsule at a 6 μm voxel size. The scans focused on a 6 mm diameter by 6 mm tall sub-region (or ROI as shown in Figure 4) of the powder capsule.

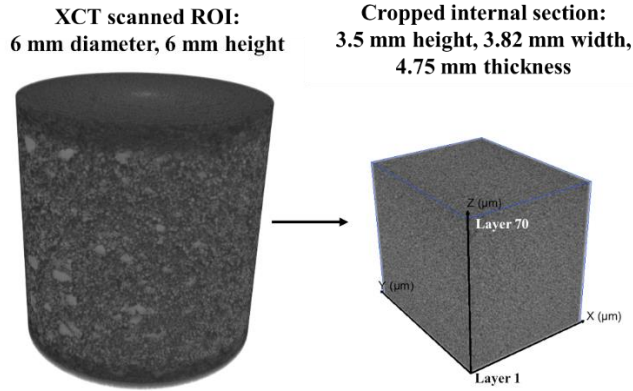

Figure 4 Depiction of the cropped internal section from the original XCT scanned ROI

The CT scans were reconstructed using the ZEISS Scout-and-Scan™ Control System Reconstruction Software package to produce a series of gray-scale images with 16-bit intensity ranges as shown Figure 5(a). The image stack was then cropped in ImageJ to obtain a rectangular central region of length ($Z$): 3.50 mm, width ($X$): 3.82 mm, thickness ($Y$): 4.75 mm (as shown in Figure 4) of the internal section of the sample. The printing layer thickness was set to 50 μm and therefore the 3.5 mm translates to 70 printing layers. The cropping was done so as to only capture the powder cake during the steady-state part of manufacturing, when the capsule internal walls are vertical and away from the top and bottom edges, which may add uncertainty due to the different heating regime at those locations.

## 2.4 Analysis of XCT data to extract particle coordination number, packing density, contact size ratio and sinter neck size

Noise was reduced by down-sampling the gray-scale images to 8-bit in MATLAB. A gray-scale threshold value to isolate the solid portion of the specimen was chosen such that the thresholded images had a qualitative visual match to the gray-scale data pore structure. After down-sampling and thresholding, a black and white or B&W image (white = solid, black = pore) stack was obtained from MATLAB (Figure 5(b)). MATLAB code for down-sampling, thresholding and obtaining the B&W image can be made available upon request. This B&W image stack obtained from MATLAB was then used as an input for subsequent processing in Python. PoreSpy, a Python Toolkit for quantitative analysis of porous media images, developed by Gostick et al. [40] was used for 3D analysis of the image stack. As a first step, PoreSpy performed a connected component analysis on the B&W image stack to obtain the layer-wise density by comparing the ratio of black and white voxels contained within the volume for each layer for the entire sample. Such a method for obtaining the layer-wise density was employed previously by the authors in [41]. As a second step, a water shed segmentation was performed to isolate and label every particle. Detailed information on the watershed algorithm can be found in [42], [43]. Thirdly, particles in contact with the image border were removed. The removal of border particles was done to reduce any quantification errors in subsequent analysis. Lastly, interface areas (connected areas between identified particles) of the sintered regions were obtained.

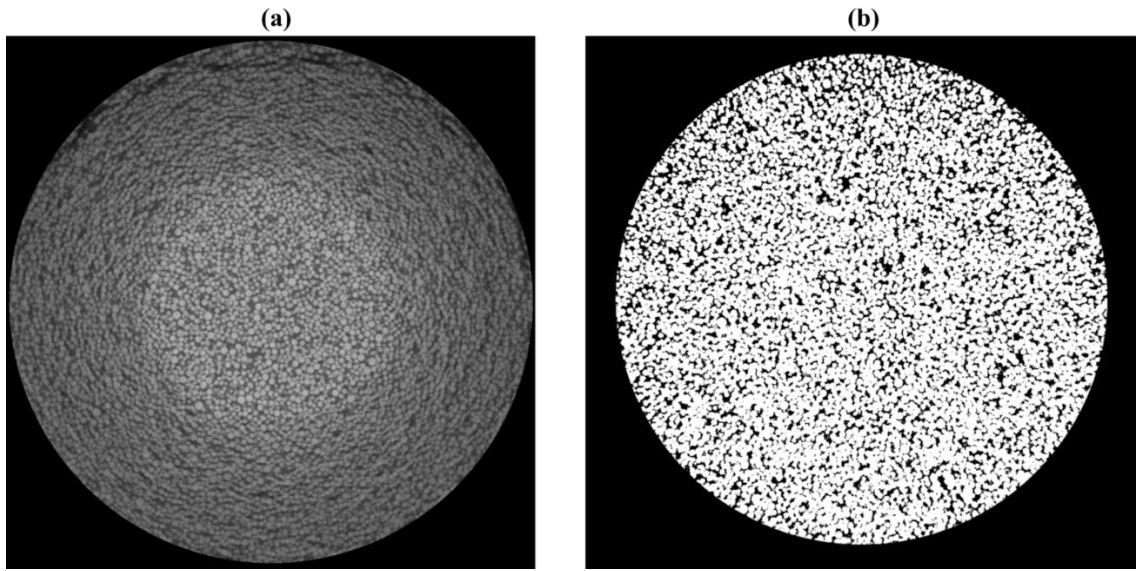

Figure 5 (a) Grayscale image of a slice of powder artefact obtained by XCT reconstruction (b) Segmented image of a slice obtained by employing thresholding in MATLAB

The particle space extracted based on the methodologies described above was used to extract the average particle coordination number (i.e., total number of particles that are in contact with an individual particle), the contact area between powder particles ($\mu m^2$), the layer-wise powder bed density (%), the equivalent diameter of the powder particles ($\mu m$), the X-, Y- and Z- coordinates of the powder particle centroids ($\mu m$), the sinter neck area between powder particles ($\mu m^2$), the sinter neck diameter between powder particles ($\mu m$), and sinter neck diameter to particle diameter ratio; the Python code used for processing can be provided upon request.

## 2.5 Inferred effective thermal conductivity

The thermal conductivity can be calculated based on the methodology described by Gusarov et al. [25], [44], where they define the contact effective thermal conductivity between particles, $\lambda_{eff}$, as:

$$\frac{\lambda eff}{\lambda} = \frac{pn}{\pi} x \qquad \qquad 3$$

where $\lambda$ is the thermal conductivity of the corresponding bulk material (for Ti6Al4V the value is 6.7 W/m/K as per [45]), $p$ is the packing density of the powder layer, $n$ is the coordination number, $x$ is the contact size ratio The contact size ratio is defined as the ratio of the contact spot diameter (or sinter neck diameter) to the sphere diameter (or powder particle diameter). The effective thermal conductivity (W m$^{-1}$ K$^{-1}$) of every powder layer of the capsule inside the powder capture artefact was then obtained from Python, as per Equation 3.

## 3 Results and Discussion

According to Gusarov et al. [25], [44], the effective thermal conductivity of a randomly packed powder bed depends on three in-situ properties: the packing density, the contact size ratio and the coordination number. In this study, it is shown that these in-situ properties of the powder cake, in an EB-PBF machine, can be measured by using a special powder capsule inside a powder-capture artefact (as shown in Figure 2), trapping the powder cake inside of it. For simplicity, three samples were selected – one each from the top, middle and bottom of the build plate; and data pertaining to these is shown in the subsequent sections. Data related to all other samples can be found in the supplementary data document.

### *3.1* **Packing density of powder layer, *p***

Figure 6 depicts the packing density of the powder layer (also known as layer-wise density) v/s layer number for the LUC, C and RUC samples for all experiments mentioned in Table 2. The data for the remaining samples is presented in Figure S1 of the supplementary data document. The average layer-wise density values lie between 58.42 ± 1.10 % and 59.57 ± 1.48 % for all Experiment 1 powder capsules (i.e., preheating temperature of 650 °C), 59.40 ± 1.70 % and 60.54 ± 1.93 % for all Experiment 2 powder capsules (i.e., preheating temperature of 690 °C) and 60.83 ± 0.99 % and 61.87 ± 0.96 % for all Experiment 3 powder capsules (i.e., preheating temperature of 730 °C) as shown in Table 3.

Firstly, it can be observed that with an increase in preheating temperature, the layer-wise density values increase. Secondly, the densities were predominantly uniform across the various locations sampled on the build platform, for a given preheating temperature. Thirdly, there is a clear indication that the layer-wise density for any specific capsule, across the 70 layers, had a global variation of approximately ± 1.5% with a clear density increase in the central region of the capsules. The reason for the density increase in the central region of the capsules is unclear at the moment but may be affected by the heat transfer phenomena at the bottom and top of the capsule. The cause of such outcomes will be examined in a follow up study.

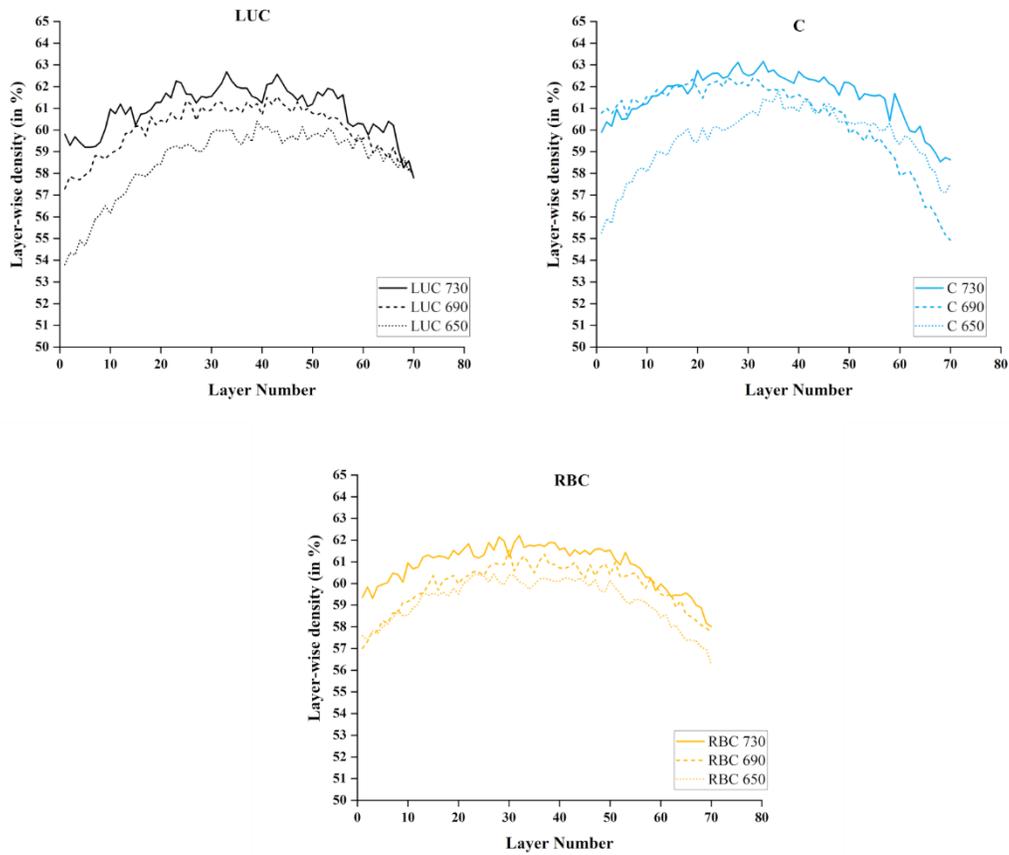

Figure 6 Layer-wise density (in %) v/s Layer number for LUC, C and RBC samples. The solid lines, dashed lines and dotted lines represent the behaviour of the capsules processed at 730 °C, 690 °C, and 650 °C, respectively.

In the EB-PBF process, a defocused electron beam is used as a planar heat source to preheat the build plate to a desired sintering temperature and control the heat input and sintering mechanism. A few studies in literature have looked at influencing the sintering mechanism of the powder bed by changing the PH1 line energy [13], PH2 line energy [8], beam current [39], etc. Leung et al. [13] and Smith et al. [8] observed a slight increase in the packing density of the powder cake with an increase in the PH1 and PH2 line energy, respectively. However, the current study is one of the first studies that looks at influencing the heat input, sintering, and consequently the density of each powder layer by, controlling the preheating temperature. Powder particles that are in contact with each other start forming necks at the contact points. With an increase in the preheating temperature, the neck size increases. This leads to further densification and reduction in the porosity. This also leads to an increase in the coordination number as seen in Table 3.

Table 3 Table summarizing the average values and standard deviation for packing density, contact size ratio, eff. thermal conductivity and particle equivalent diameter, for all powder capsules

| | Packing density, $p$ (%) | | | Contact size ratio, $x$ | | | Effective thermal conductivity, $\lambda_{eff}$ (W/m/K) | | | Particle equivalent diameter (µm) | | | Average Coordination Number, n | | |
|---|---|---|---|---|---|---|---|---|---|---|---|---|---|---|---|
| Preheat temp. (°C) | 650 | 690 | 730 | 650 | 690 | 730 | 650 | 690 | 730 | 650 | 690 | 730 | 650 | 690 | 730 |
| Exp. No. | 1 | 2 | 3 | 1 | 2 | 3 | 1 | 2 | 3 | 1 | 2 | 3 | 1 | 2 | 3 |
| LUC | 58.56 ± 1.63 | 60.01 ± 1.13 | 60.91 ± 1.09 | 0.44 ± 0.008 | 0.47 ± 0.004 | 0.47 ± 0.004 | 1.76 ± 0.11 | 1.94 ± 0.07 | 2.02 ± 0.07 | 54.51 ± 0.72 | 55.51 ± 0.55 | 56.44 ± 0.56 | 3.39 ± 0.08 | 3.49 ± 0.05 | 3.52 ± 0.05 |
| LMC | 59.45 ± 1.14 | 60.27 ± 1.16 | 61.87 ± 0.95 | 0.45 ± 0.004 | 0.47 ± 0.004 | 0.48 ± 0.003 | 1.84 ± 0.07 | 1.96 ± 0.08 | 2.11 ± 0.06 | 55.56 ± 0.60 | 55.48 ± 0.60 | 56.04 ± 0.50 | 3.43 ± 0.05 | 3.50 ± 0.05 | 3.58 ± 0.05 |
| LBC | 58.42 ± 1.10 | 59.40 ± 1.70 | 61.14 ± 0.98 | 0.45 ± 0.004 | 0.46 ± 0.010 | 0.48 ± 0.003 | 1.74 ± 0.06 | 1.89 ± 0.12 | 2.04 ± 0.06 | 56.64 ± 0.56 | 56.13 ± 0.73 | 56.22 ± 0.54 | 3.36 ± 0.05 | 3.44 ± 0.08 | 3.53 ± 0.04 |
| TM | 59.19 ± 1.44 | 60.13 ± 1.14 | 61.32 ± 0.96 | 0.45 ± 0.007 | 0.47 ± 0.004 | 0.48 ± 0.003 | 1.82 ± 0.10 | 1.95 ± 0.07 | 2.05 ± 0.06 | 54.03 ± 0.67 | 55.16 ± 0.53 | 56.77 ± 0.48 | 3.45 ± 0.07 | 3.50 ± 0.05 | 3.54 ± 0.05 |
| C | 59.57 ± 1.47 | 60.53 ± 1.93 | 61.55 ± 1.15 | 0.46 ± 0.005 | 0.47 ± 0.012 | 0.48 ± 0.004 | 1.87 ± 0.10 | 1.99 ± 0.15 | 2.08 ± 0.08 | 55.63 ± 0.57 | 55.47 ± 0.75 | 56.29 ± 0.55 | 3.44 ± 0.08 | 3.53 ± 0.09 | 3.57 ± 0.05 |
| BM | 58.69 ± 1.80 | 59.97 ± 1.13 | 61.29 ± 1.32 | 0.45 ± 0.011 | 0.47 ± 0.005 | 0.48 ± 0.006 | 1.77 ± 0.13 | 1.93 ± 0.07 | 2.05 ± 0.10 | 56.76 ± 0.81 | 56.77 ± 0.57 | 56.84 ± 0.53 | 3.37 ± 0.09 | 3.46 ± 0.05 | 3.54 ± 0.07 |
| RUC | 59.31 ± 1.11 | 59.62 ± 1.27 | 60.97 ± 1.00 | 0.45 ± 0.004 | 0.46 ± 0.006 | 0.47 ± 0.004 | 1.83 ± 0.06 | 1.88 ± 0.08 | 2.03 ± 0.06 | 54.94 ± 0.56 | 54.95 ± 0.62 | 56.24 ± 0.51 | 3.44 ± 0.05 | 3.46 ± 0.05 | 3.54 ± 0.05 |
| RMC | 58.94 ± 1.37 | 60.37 ± 1.39 | 61.65 ± 1.34 | 0.45 ± 0.007 | 0.47 ± 0.007 | 0.47 ± 0.006 | 1.80 ± 0.09 | 1.97 ± 0.10 | 2.06 ± 0.10 | 54.82 ± 0.65 | 55.58 ± 0.60 | 56.30 ± 0.65 | 3.42 ± 0.06 | 3.52 ± 0.06 | 3.57 ± 0.07 |
| RBC | 59.20 ± 1.05 | 59.88 ± 1.09 | 60.82 ± 0.98 | 0.45 ± 0.004 | 0.47 ± 0.004 | 0.47 ± 0.005 | 1.82 ± 0.06 | 1.92 ± 0.07 | 1.99 ± 0.06 | 57.04 ± 0.51 | 56.68 ± 0.62 | 56.62 ± 0.53 | 3.40 ± 0.05 | 3.46 ± 0.05 | 3.50 ± 0.04 |

It is important to evaluate whether the change in the preheating temperature also causes a change in the density of the solid part. Therefore, specimen C from Experiment 1, Experiment 2 and Experiment 3 was taken as a representative sample. A small section of the solid part, of the powder-capture artefact, was cropped (Figure 7) and its layer-wise density was evaluated.

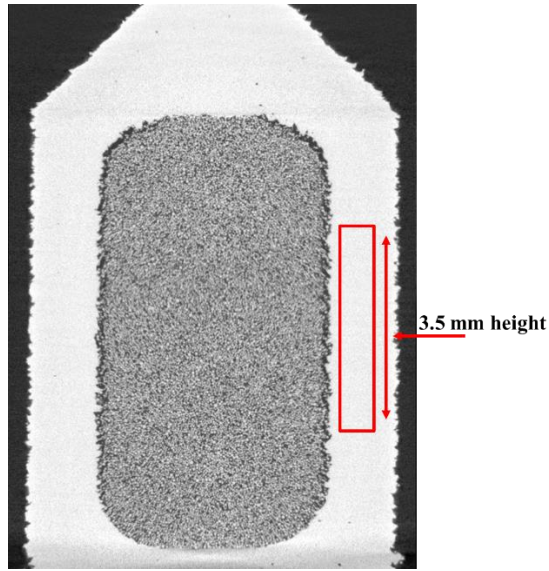

Figure 7 Depiction of the solid part of the powder-capture artefact that was cropped and analyzed to obtain layer-wise density

It can be observed from Figure 8 that the layer-wise density for all three samples is quite consistent. The average layer-wise density values for the Experiment 1, Experiment 2, and Experiment 3 samples are 99.98%, 99.97% and 99.98% respectively. Hence it can be concluded that the change in preheating temperatures does not lead to a change in the density of the solid parts built in the powder cake.

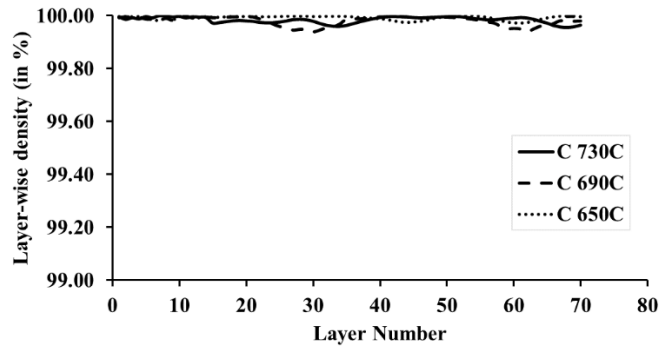

Figure 8 Layer-wise density (in %) v/s Layer number C samples. The solid lines, dashed lines and dotted lines represent the behaviour of the solid part, of the powder-capture artefact, processed at 730 °C, 690 °C, and 650 °C, respectively.

It is important to assess the impact of lower preheating temperatures since high preheating temperatures can cause excessive partial melting which may reduce the reusability of the powder [46], make de-powdering difficult [8], and lead to other disadvantages as described by Sigl et al. [12]. Furthermore, Suard et al. [47] and Koike et al. [48] demonstrated that excessive partial melting leads to high surface roughness containing rippled structures and visible sintered powder grains, in EB-PBF. Here, it can be noted that a decrease 80 °C in the preheating temperature led to a mere approximate decrease of 3% in the relative layer-wise density of the powder-cake and no change in density of the solid part.

### 3.2 *Average Equivalent diameter of powder particles in the powder cake*

The particle size has a significant influence on the layer-wise powder bed density [49], [50] and furthermore on the thermal conductivity. Alkahari et al. [33] showed that, for their LPBF experiments, the thermal conductivity increased with an increase in powder particle diameter. This behaviour was attributed to the conductive heat transfer mechanism between particles. They mentioned that during conduction from the heat source to the thermocouple measurement point, smaller powder particles experienced increased repetitive changes between solid and air as a medium, however, the larger particles behaved the other way around. As a result, heat transfer and consecutively the thermal conductivity, is higher for the larger powder particles [33]. Therefore, it is imperative to observe whether there is a change in the powder particle equivalent diameter with a change in the preheating temperature. This will help isolate all factors that cause changes (if any) in the effective thermal conductivity. In order to calculate the particle size, all particles were assumed to be completely spherical. Figure 9 depicts the layer-wise average equivalent powder particle diameter (in µm) v/s layer number for the LUC, C and RUC samples for all experiments. The data for all remaining samples is provided in Figure S2 of the supplementary data document.

It is observed that the average equivalent diameter values lie between 54.03 ± 0.68 µm to 57.05 ± 0.52 µm for all Experiment 1 powder capsules, 54.95 ± 0.62 µm to 56.77 ± 0.58 µm for all Experiment 2 powder capsules and 56.04 ± 0.51 µm to 56.84 ± 0.53 µm for all Experiment 3 powder capsules (as shown in Table 3). From Table 3 and Figure 9, it can be stated that the average equivalent diameter is quite consistent for all preheating temperatures and across all capsule locations.

Therefore, a change in the layer-wise density or thermal conductivity associated with the preheating temperature cannot be correlated to the powder particle equivalent diameter. Another observation from Figure 9 is that there is an increase in the average equivalent diameter in the central region of the capsules. Upon close observation we can see that the global variation across the 70 layers (Table 3) is approximately less than 1%. Such a variation is considered to be negligible for the current study.

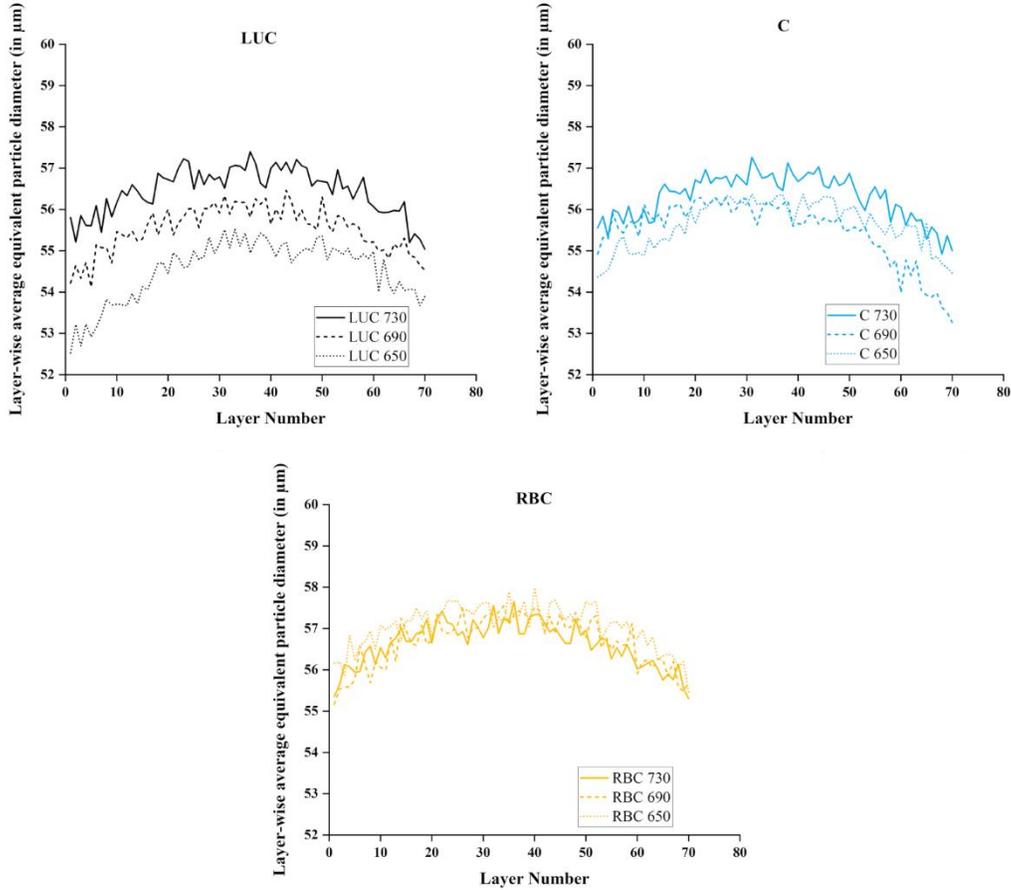

Figure 9 Layer-wise average equivalent powder particle diameter (in µm) v/s Layer number for LUC, C and RBC samples. The solid lines, dashed lines and dotted lines represent the behaviour of the capsules processed at 730 °C, 690 °C, and 650 °C, respectively.

### 3.3 *Contact size ratio, x*

The contact size ratio, *x*, is the ratio of the contact spot diameter (or sinter neck diameter) to the sphere diameter (or powder particle diameter). Obtaining this ratio is critical to identify the effective thermal conductivity according to the equation presented by Gusarov et al. [25], [44]. The contact size ratio shows how strongly the powder particles are sintered together. A higher contact size ratio value translates to better sintering. Consequently, a higher degree of sintering leads to a lower chance of repulsion by virtue of electrostatic charging [3]. Figure 10 depicts the layer-wise average contact size ratio v/s Layer number for the LUC, C and RUC samples for all experiments. The data for the remaining samples is presented Figure S3 of the supplementary data document.

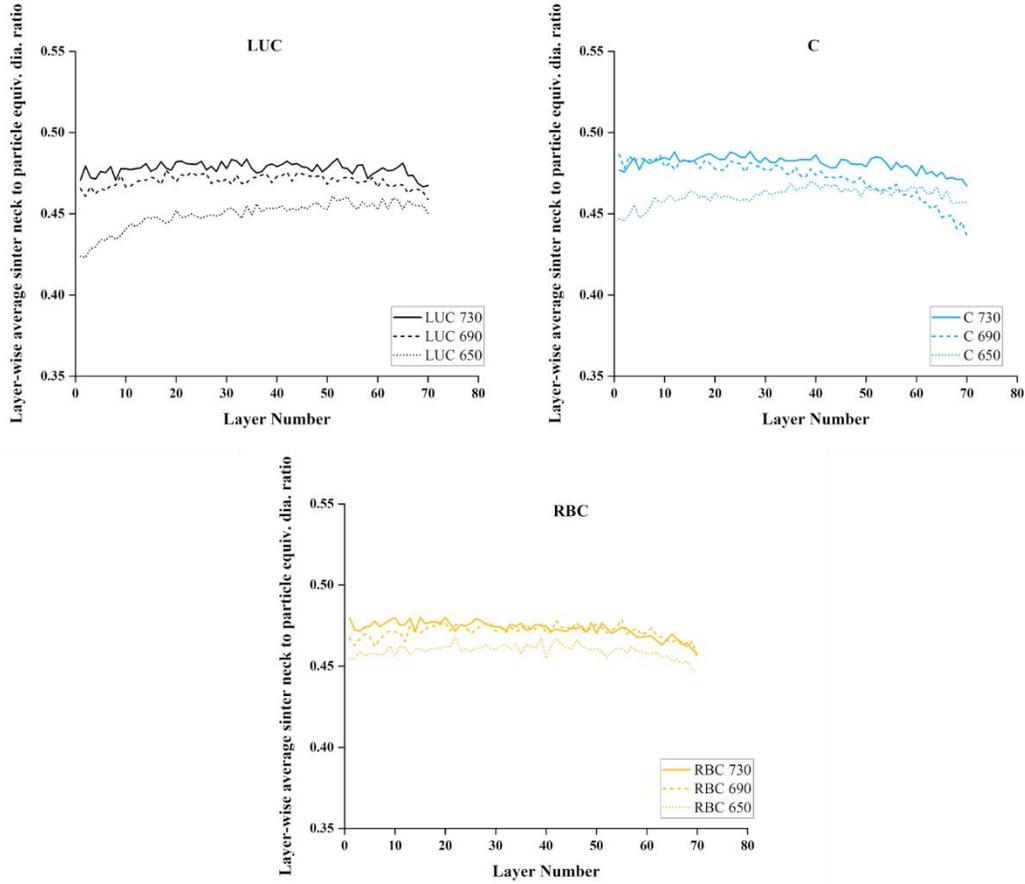

Figure 10 Layer-wise average sinter neck diameter to particle equivalent diameter ratio v/s Layer number for LUC, C and RBC samples. The solid lines, dashed lines and dotted lines represent the behaviour of the capsules processed at 730 °C, 690 °C, and 650 °C, respectively.

It is observed that the average contact size ratio values lie between 0.45 ± 0.004 to 0.46 ± 0.005 for all Experiment 1 capsules, 0.46 ± 0.006 to 0.47 ± 0.005 for all Experiment 2 capsules, and 0.47 ± 0.005 to 0.48 ± 0.003 for all Experiment 3 capsules (as shown in Table 3). It is observed that there is an increasing contact size ratio response as a function of increasing the preheating temperature. It is also observed that the contact size ratios were mostly uniform across the various locations sampled on the build platform, for a given preheating temperature.

In Figure 9 it is noted that the equivalent diameter range is quite consistent for all preheating temperatures. This means that the only variable changing in the contact size ratio is the sinter neck diameter. A previous study performed on assessing the powder sintering mechanism of EB-PBF powder beds [51] concluded that solid-state sintering is the dominant mechanism during the preheating procedure. In addition to experimental studies from other authors [5], [6], [13], they also concluded that sintering of powder particles in EB-PBF Ti-6Al-4V only occurs between 873 K (600 °C) and 1003 K (730 °C). No true sintering occurs below 873 K [5], [51]. They also demonstrated that sinter necks grow faster with an increase in preheating temperatures. This is consistent with the current study where an increase in the contact size ratio is observed with an increase in preheating temperatures. Specifically, an increase of 80 °C in the preheating temperature led to an approximate increase of 5% in the contact size ratio. This increase in the sinter neck diameter can be attributed to the increase in layer-wise density with preheating temperature. As the density increases, the contact points between powder particles improve as more powder particles are touching their adjacent particles. As explained by Wheat et al. [50] there are two different types of sintering mechanisms within solid-state sintering: non-densifying and densifying sintering mechanisms (as shown in Figure 11). The authors hypothesize that the solid-state sintering mechanism taking place in EB-PBF is the densifying mechanism where mass from the particles moves from the core to the neck of the particles [52]. Furthermore, understanding which mechanism (lattice diffusion, grain boundary diffusion or plastic flow) is causing densification is beyond the scope of the current study.

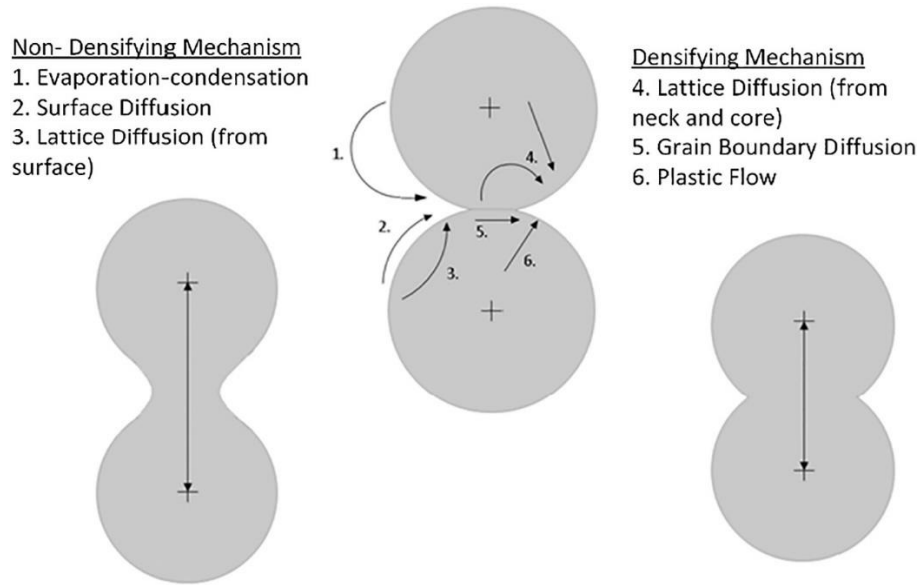

Figure 11 Possible solid-state sintering mechanisms: non-densifying mechanisms (left) and densifying mechanisms (right). Reprinted with permission from Wheat et al. [50]

To assess the correlation between density, coordination number, and contact size ratio; the values for a given experiment were averaged and plotted in Figure 12. The coefficient of determination (or R-squared) values depict that a near linear relationship exists between the density, coordination number, and the contact size ratio.

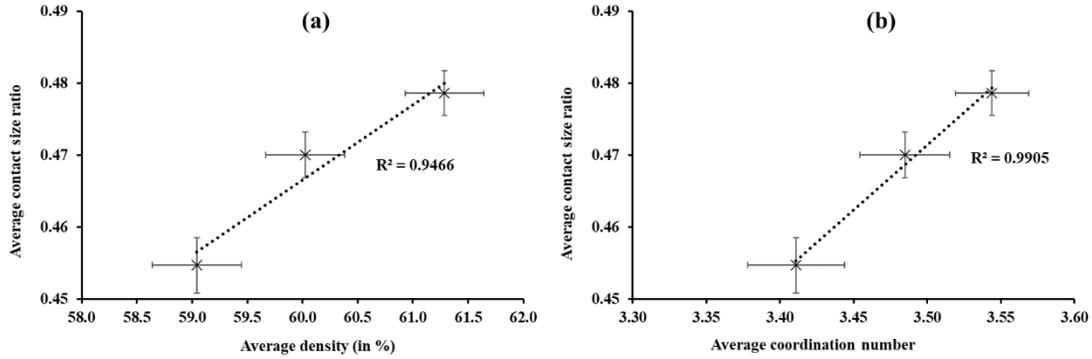

Figure 12 Average (a) density (b) Coordination number v/s average contact size ratio for all experiments with linear trendlines and R-squared values. Error bars represent the standard deviation.

### 3.4 Effective thermal conductivity, $\lambda_{eff}$

Thermal conductivity is a measure of the rate of heat transfer through material [33]. The effective thermal conductivity, $\lambda_{eff}$, was calculated for every sinter neck in the powder cake. However, the layer-wise average effective thermal conductivity of LUC, C, and RBC samples, was used for demonstration in Figure 13. The data for the remaining samples is presented in Figure S4 of the supplementary data document.

It is observed that the average effective thermal conductivity values lie between 1.75 ± 0.07 W/m/K to 1.87 ± 0.10 W/m/K for Experiment 1 capsules, 1.88 ± 0.09 W/m/K to 1.99 ± 0.15 W/m/K for all experiment 2 capsules and 1.99 ± 0.07 W/m/K to 2.11 ± 0.07 W/m/K for all Experiment 3 capsules (as shown in Table 3). An increase in preheating temperature leads to an increase in the effective thermal conductivity of the powder cake. Specifically, an increase of 80 °C in the preheating temperature led to an approximate increase of 12% in the effective thermal conductivity.

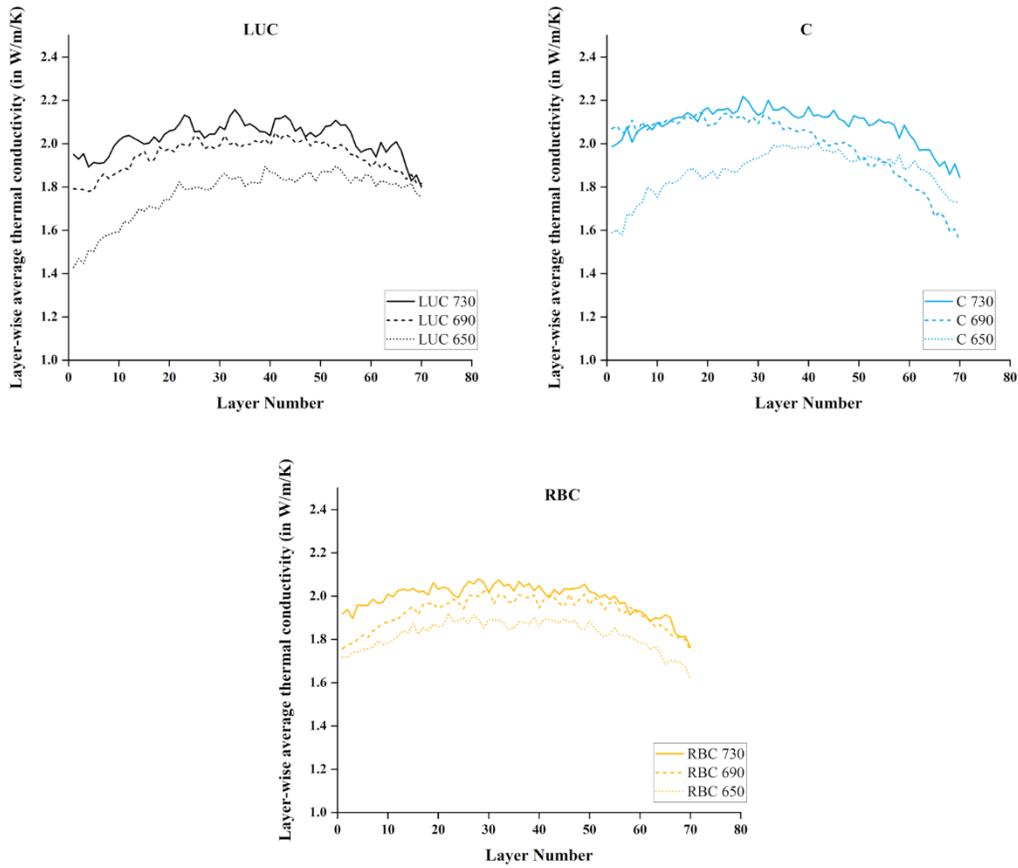

Figure 13 Layer-wise average effective thermal conductivity v/s Layer number for LUC, C and RBC samples. The solid lines, dashed lines and dotted lines represent the behaviour of the capsules processed at 730 °C, 690 °C, and 650 °C, respectively.

To assess the correlation between density, contact size ratio, coordination number, and the effective thermal conductivity; the values for a given experiment were averaged and plotted in Figure 14. The R-squared values depict that a near linear relationship exists for all the metrics. It is a well-known fact that the thermal conductivity of packed powder beds increases with temperature [53]–[55], however, the unanswered question is how well are these two metrics correlated.

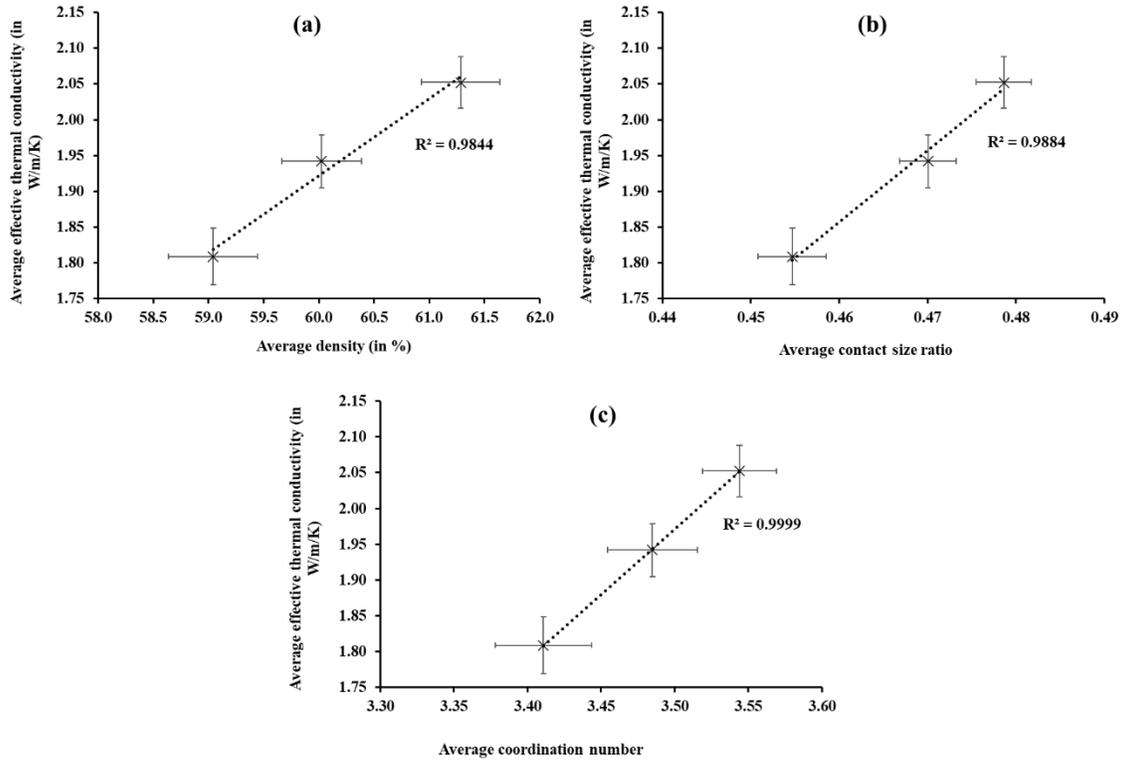

Figure 14 Average (a) density (b) contact size ratio (c) coordination number v/s average effective thermal conductivity for all experiments with linear trendlines and R-squared values. Error bars represent the standard deviation.

## 3.5 Correlation between the preheating temperatures and density, contact size ratio and effective thermal conductivity of the powder capsules, respectively

Stemming from the observations in Sections 3.1 - 3.4, some correlations were noticed between the preheating temperature and the various metrics such as layer-wise density, contact size ratio, coordination number and effective thermal conductivity. Therefore, to quantify this interdependency, the packing density, contact size ratio, coordination number, and effective thermal conductivity values, as shown in Table 3, for a given experiment were averaged and plotted against the preheating temperatures in Figure 15. The R-squared values depict that a near linear relationship exists between the preheating temperatures and the various metrics.

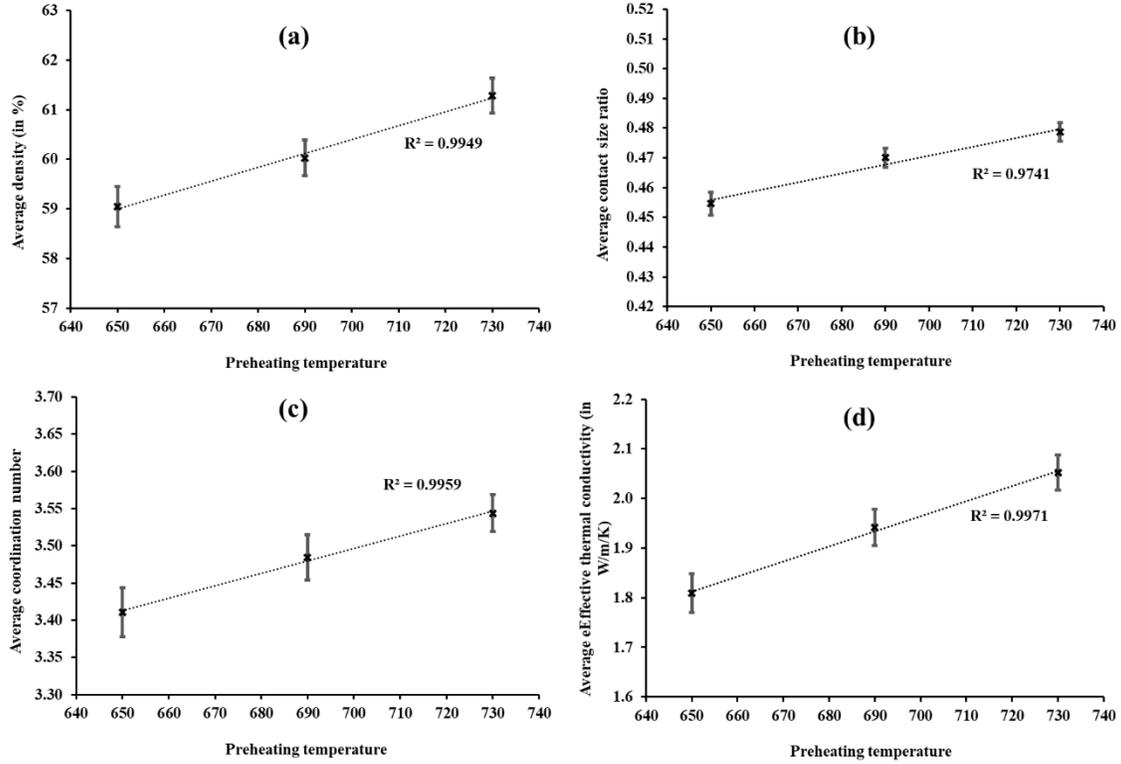

Figure 15 Average (a) density, (b) contact size ratio, (c) coordination number, and (d) effective thermal conductivity, v/s preheating temperature for all experiments with linear trendlines and R-squared values. Error bars represent the standard deviation.

A linear equation was established from the experimentally-measured average values of the packing density (Equation 4), the contact size ratio (Equation 5), coordination number (Equation 6), and the effective thermal conductivity (Equation 7) for the three preheating temperatures.

$$p = 0.028 \times T + 40.784 \qquad 4$$

$$x = 0.0003 \times T + 0.2613 \qquad 5$$

$$n = 0.0017 \times T + 2.3318 \qquad 6$$

$$\lambda eff = 0.003 \times T - 0.1648 \qquad 7$$

where $p$ is the packing density, $x$ is the contact size ratio defined as the ratio of the contact spot diameter (or sinter neck diameter) to the sphere diameter (or powder particle diameter), n is the coordination number, $\lambda eff$ is the effective thermal conductivity, and $T$ is the preheating temperature. These equations can be used to predict the density, contact size ratio, coordination number, and effective thermal conductivity of the powder bed when using preheating temperatures between 650 °C and 730 °C.

Maly et al. [56] have demonstrated that an increase in preheating temperature in Ti-6Al-4V samples, produced by LPBF, led to dramatic degradation of powder. An increase in the $O_2$ and $H_2$ content, beyond the ASTM defined limit, was observed. Al-Bermani et al. [39] have demonstrated that the yield strength and tensile strength decrease with an increase in the processing temperatures. Therefore, low preheating temperatures should be used for partially sintering the powder bed in EB-PBF. The current study shows that it is possible to use a reduced preheating temperature with very little change in the layer-wise density and effective thermal conductivity of the powder cake. Furthermore, the empirically-derived model for predicting the density, contact size ratio, coordination number, and effective thermal conductivity can be used to study the necking and sintering phenomenon for use in FE models. The authors believe that this work will aid in providing reliable empirical information into furthering the understanding behind the evolution of the input Ti-6Al-4V powder from its as-received powder feedstock state to preheated, partially sintered molten and finally its solid state.

# 4 Conclusions

The effect of varying preheating temperatures on the effective thermal conductivity of the Ti-6Al-4V powder cake, in EB-PBF, were observed. Some of the main findings were:

i. The calculated average packing density, contact size ratio, coordination number, and effective thermal conductivity, of the powder-capsules, ranged from $58.42 \pm 1.10$ % to $61.87 \pm 0.96$ %, $0.45 \pm 0.004$ to $0.48 \pm 0.003$, $3.36 \pm 0.05$ to $3.58 \pm 0.05$, and $1.75 \pm 0.07$ W/m/K to $2.11 \pm 0.07$ W/m/K, respectively when the preheating temperature was varied between 650 °C and 730 °C.
ii. The effective thermal conductivity of the powder cake at a given preheating temperature strongly depends on the packing density, contact size ratio and coordination number.
iii. The current study shows that it is possible to use a reduced preheating temperature since a mere change of 3% in the layer-wise density and 12% in the effective thermal conductivity, was observed, with an 80 °C decrease in the preheating temperature.
iv. A decrease in the preheating temperature led to a linear decrease in the packing density, contact size ratio, coordination number, and consequently the effective thermal conductivity. Logarithmic regression equations were established from the empirically - derived thermal conductivity data. These equations can be used to predict the thermal conductivity of the powder cake, in EB-PBF, when using preheating temperatures between 650 °C and 730 °C.

# 5 Funding

The authors appreciate the partial funding support received from the Natural Sciences and Engineering Research Council (NSERC) Holistic Innovation in Additive Manufacturing (HI-AM) grant NETGP 494158 - 16.

# 6 Conflicts of Interest

The authors declare no conflict of interest.

Binder Jetting Additive Manufacturing," *J. Manuf. Sci. Eng.*, vol. 142, no. 10, Oct. 2020, doi: 10.1115/1.4047140.